\documentclass{ws-procs975x65}
\usepackage[T1]{fontenc}
\usepackage{amssymb}
\usepackage[arrow,matrix]{xy}

\newcommand{\C}{\mathbb{C}}
\newcommand{\cH}{\mathcal{H}}
\newcommand{\act}{\triangleright}
\newcommand{\ares}{\bar{a}}
\newcommand{\Obs}{\mathcal{O}}

\title{Renormalization for spin foam models\\ of quantum gravity}
\author{Robert Oeckl\footnote{email: oeckl@cpt.univ-mrs.fr}}

\address{Centre de Physique Th\'eorique,\\
CNRS Luminy, 13288 Marseille, France}

\begin{document}
\maketitle

\abstracts{We give an introductory account to the renormalization of
  models without metric background. We sketch the application to
  certain discrete models of quantum gravity such as spin foam models.}

\section{Conventional Renormalization}

We first describe the conventional renormalization procedure as it
applies to lattice models of statistical field theory (see e.g.\
\cite{ItDr:statftheo}). We do this in a
way that is suitable for the generalization performed later. To be
specific, the renormalization prescription we consider uses a
parameter space fixed from the outset rather than a decimation
prescription. A prime example where this type of renormalization is of
major importance is lattice gauge theory.

Consider a lattice model of a statistical field theory. Local
degrees of freedom (e.g.\ spins, group elements) are associated with
elements of the lattice (vertices, links, plaquettes, etc.), which
represents a discretization of space or space-time. Physical
quantities (energy, correlation functions, etc.) are obtained by
summing over these degrees of freedom with certain weights. These
quantities depend on the discretization scale (lattice spacing) $a$
and parameters $\lambda$ of the model, often called \emph{coupling
  constants}. We will denote such quantities
generically by $\Obs(a,\lambda)$.

The real physical system one is trying to model might be a continuous
system (such as a quantum field theory). In this case the lattice
model approximates the physical system for small lattice spacings
$a$. An example for this situation is lattice
gauge theory. On the other hand the physical system might
have a lattice like structure, but with a lattice spacing different
from $a$. In this case the model might be intended to give an
effective description of the physical system. This is typically the
case in condensed matter physics. In both cases, the lattice spacing
$a$ is a mathematical artifact. To extract physical predictions from
the model one has to \emph{renormalize}. Very broadly speaking this
means having control over the dependence of physical quantities on the
discretization scale $a$.

We formalize this as follows. We denote by $A$ the set of
discretizations (lattices). A given discretization $a$ is determined
completely by its lattice spacing, which we denote by the same symbol
$a$. To change $a$ to $a'$ we simply
multiply by a positive real number $g$. These form the group $G$ of
scale transformations, which we also call the group of changes of
discretizations.

The model comes with coupling constants (parameters) $\lambda$ which
are adjustable from within a space of coupling constants
$\Lambda$. The renormalization problem might now be described as
follows: Find an action of the group $G$ of changes of
discretizations on the space $\Lambda$ of coupling constants such that
physical parameters of the model remain unchanged.
Formally, this reads
\begin{equation}
\Obs(g\act a, g\act \lambda) = \Obs(a,\lambda),
 \qquad \forall \Obs, \forall a\in A, \forall g\in G, \forall
 \lambda\in \Lambda.
\label{eq:reneq}
\end{equation}
The symbol $\act$ stands for the action. In particular, $g\act a$ is
just the multiplication of $a$ by $g$.

Given a solution of equation (\ref{eq:reneq}) one calls $G$ (with its
actions) the \emph{renormalization group} and the orbits of $G$ in $\Lambda$
are called the \emph{renormalization group flow}. This renders the
model physically predictive in the following sense: For a given orbit,
the quantities $\Obs$ of the model take definite values
independent of the discretization.

For brevity we have simplified the situation considerably,
leaving out significant details. The equation
(\ref{eq:reneq}) usually cannot be required to hold exactly, but only
approximately. The degree of approximation is normally linked to the
scale $a$. Furthermore, physical quantities $\Obs$ often cannot
be defined for arbitrary $a$, but usually require $a$ to be sufficiently
small compared to some physical scale (e.g.\ for correlation
functions).

\section{Renormalization without background}
\label{sec:renfree}

We now proceed to generalize the renormalization idea to discrete
theories without metric background \cite{Oe:renormdisc}. These are
models which use a discretization of space (or space-time), where no
metric structure is provided and thus no notion of scale
exists. Consequently, the notion of regular discretization (as e.g.\ a
hypercubic lattice) makes no sense. Such discretizations can be for
example triangulations of a manifold, or CW-complexes, or even abstract
graphs. Again, degrees of freedom (spins, group elements etc.) are
associated to elements of the discretization (vertices, cells, etc.). 
Physical quantities $\Obs$ are obtained by a weighted sum over
the degrees of freedom.

As in the familiar lattice setting the physical system one tries to
model might be continuous or discrete. Irrespective of this the
principal idea of renormalization remains the same as before: Control
the change of the model with change of discretization to extract physical
predictions.

The model
in question of course needs to have adjustable coupling constants as in the
lattice case. However, in general this will not be enough. A
discretization and thus also a change of discretization is something
much more general than in the lattice case. There a change of
discretization was a \emph{global} operation, necessarily affecting
the whole lattice by rescaling it. Now a change of discretization is
much more arbitrary. In particular, we may change a discretization
\emph{locally}, i.e.\ change only a little part of it. To
effectively compensate such a local change by tuning coupling
constants is in general only conceivable if these coupling constants
are local. That is, we require coupling constants also to be attached
to elements of the discretization.

We now follow essentially the same route as in the lattice setting,
introducing modifications on the way as required. As before we call
the space of discretization $A$ and a discretization itself $a$. (Note
that $a$ is no longer identified with a scale or number.) What is
a change of discretization? We use the most general notion possible,
namely specify a change by a pair $(a,a')$ of initial and final
discretization. These pairs form a \emph{groupoid} $G$. As in a group
elements can be composed, but now only if they match. Matching means
here simply that the final discretization of the first pair is the
initial discretization of the second one. Concretely, $(a,a')$ composes
with $(a',a'')$ to $(a,a'')$.

The space of coupling constants is a different one for each
discretization $a$ and we denote it by $\Lambda_a$, with a choice of
coupling constants denoted by $\lambda_a$. The equivalent of the
fixed-point equation (\ref{eq:reneq}) formally takes almost the
identical form
\begin{equation}
\Obs(g\act a, g\act \lambda_a) = \Obs(a,\lambda),
 \qquad \forall \Obs, \forall a\in A, \forall g\in G_a, \forall
 \lambda_a\in \Lambda_a . 
\label{eq:reneq2}
\end{equation}
Here, $G_a$ denotes the subset of $G$ that have initial discretization
$a$. These are the elements of the form $g=(a,a')$. The crucial
ingredient is now an action of the groupoid $G$ on the spaces of
coupling constants $\Lambda$. This means the following: For each pair
of discretizations $(a,a')$ there is a map $\Lambda_a\to
\Lambda_{a'}$, sending $\lambda_a\mapsto (a,a')\act\lambda_{a'}$. These
maps are required to satisfy the composition property
$(a',a'')\act((a,a')\act\lambda_a)=(a,a'')\act\lambda_a$ and unity
property $(a,a)\act\lambda_a=\lambda_a$ in analogy to a group action.

Given a solution of equation (\ref{eq:reneq2}) we call $G$ (with its
actions) the \emph{renormalization groupoid} and the orbits of $G$ in
$\Lambda$ the \emph{renormalization groupoid flow}. Such a solution
renders the model predictive in the sense that for a given orbit the
quantities $\Obs$ take definite values independent of the
discretization. 

However, even more so than in our review of the lattice case we should
add that the situation is really more complicated than our brief
treatment allows to convey. In particular, in addition to the issues
already mentioned in the previous section novel ones specific to the
generalized notion of renormalization arise.
We refer the reader to \cite{Oe:renormdisc}.

\section{Application to models of quantum gravity}

We shall now outline the application of the renormalization procedure
of the previous section to a certain class of models of quantum
gravity.

To be specific, we follow the general boundary approach
\cite{Oe:catandclock,Oe:boundary} and assume that a model of quantum
gravity is formulated roughly as follows:
\begin{itemize}
 \item Basic objects are 4-dimensional regions $M$ of space-time
 (4-manifolds) with boundaries $\Sigma$. For simplicity we might
 restrict to $M$ having the topology of a 4-ball.
 \item To any boundary $\Sigma$ is associated a \emph{vector space of
 states} $\cH_\Sigma$.
 \item To any region $M$ is associated an \emph{amplitude map}
 $\rho_M:\cH_\Sigma\to\C$.
\end{itemize}
These structures are further required to satisfy the axioms of a
topological quantum field theory.
Note that $\cH_\Sigma$ is not the usual Hilbert space of quantum
mechanics, but a kind of ``square'' of it. On the other hand, physical
probabilities emerge as in standard quantum mechanics as moduli
squared of amplitudes. For details see
\cite{Oe:catandclock,Oe:boundary}.

The class of models we are interested in here can be described as
follows. For a region $M$ of space-time consider a discretization $a$,
e.g.\ a decomposition of $M$ into cells that all have the topology of a
4-ball. This discretization includes in particular a discretization $\ares$
of the boundary $\Sigma$ of $M$ (into cells that are 3-balls). With
$\Sigma$ and $\ares$ we
associate a state space $\cH_{\Sigma,\ares}$. This should be thought
of as a restricted version of the state space $\cH_\Sigma$ that we
really want to associate with the boundary according to the above
prescription. Consider another discretization $\ares'$ of $\Sigma$
that is finer than $\ares$, i.e.\ it arises by further subdividing
$\Sigma$. Then we require that there is an embedding map
$I_{\ares,\ares'}:\cH_{\Sigma,\ares}\to \cH_{\Sigma,\ares'}$.
This might be interpreted intuitively in the sense that the state
spaces $\cH_{\Sigma,\ares}$ capture more and more of a ``full''
state space $\cH_\Sigma$ the finer the discretization is.
The amplitude map of such a model depends on the discretization and takes
values on the restricted state spaces,
$\rho_{M,a,\lambda_a}:\cH_{\Sigma,\ares}\to\C$. It depends on local
coupling constants $\lambda_a$.

Models of this type can be obtained from path-integral quantization
approaches to gravity. Prominent among these are \emph{spin foam
  models}, see the review \cite{Per:sfmodels}. Very briefly, one
formulates gravity in a form that resembles a gauge theory. Then, one
discretizes space-time and defines a discrete state sum very much as
in lattice gauge theory (except that there is no regular
lattice). Finally, the group valued variables are converted to
representation (spin) valued ones by a generalized Fourier transform.
In particular, the spin foam models for (Euclidean) quantum gravity of
Reisenberger and of Barrett and Crane can be brought into
a form that almost fits our description above. What is missing are
local adjustable coupling constants. To remedy this, we recently
proposed the ``interpolating model'' \cite{Oe:renormdisc}, a tunable
model inspired by the Barrett-Crane model.

The task of renormalization is as before, to ``get rid of'' the
discretization dependence in order to make the model physically
predictive. We apply the description of
Section~\ref{sec:renfree}, adapted to the case at hand. The natural
choice for the quantities $\Obs$ are amplitudes evaluated on
states. Concretely, given a discretization $a$ of $M$, associated
coupling constants $\lambda_a$ and a state
$\psi_{\ares}\in\cH_{\Sigma,\ares}$ we set
$\Obs(a,\lambda_a)=\rho_{M,a,\lambda_a}(\psi_{\ares})$.

We have the additional complication that not only the spaces
$\Lambda_a$ of coupling constants depend on the discretization, but
also the restricted state spaces $\cH_{\Sigma,\ares}$.
Using the embedding maps the space $\cH_\Sigma$ is constructed to
``contain all'' spaces $\cH_{\Sigma,\ares}$ in the sense that the
diagrams
\[
\xymatrix{\cH_{\Sigma,\ares}\ar[d] \ar[r] & \cH_\Sigma\\
 \cH_{\Sigma,\ares'}\ar[ur]}
\]
commute. (Technically speaking $\cH_\Sigma$ is a colimit.)
Furthermore, the embedding maps can be used to define the action of a
change of discretization $g=(a,a')$ on a state
$\psi_{\ares}\in\cH_{\Sigma,\ares}$ to be $g\act
\psi_{\ares}=I_{\ares,\ares'}(\psi_{\ares})$.

The analogue of the fixed-point equation (\ref{eq:reneq2}) can now be
stated:
\begin{equation}
 \rho_{M,a,\lambda_a}(\psi_{\ares})
 =\rho_{M,g\act a,g\act \lambda_a}(g\act \psi_{\ares}) \qquad \forall
 a\in A, \forall g\in \tilde{G}_a, \forall
 \psi_{\ares}\in\cH_{\Sigma,\ares}, \forall \lambda_a\in\Lambda_a .
\label{eq:reneq3}
\end{equation}
Note that $g$ can only be a change of discretization here that is
refining (hence the tilde over $G$), see \cite{Oe:renormdisc} for more
details on this issue.

Of course, equation (\ref{eq:reneq3}) is only where the challenge
begins. To find an actual action of the renormalization groupoid that
solves the equation (approximately) is a rather non-trivial task in
general. It is in general even highly unclear if a solution exists at all.
The application of this renormalization procedure to realistic models
of quantum gravity stands only at its very beginning.

We make a final remark on spin foam models.
Usually they are not interpreted to provide a general boundary
type theory as described above. Rather, they are either interpreted as
providing transition amplitudes for a canonical quantization called
\emph{loop quantum gravity} or they are interpreted more in line with
statistical lattice models. In the latter case observable quantities
would be associated to certain weight functions measuring local
degrees of freedom in the interior of $M$. In both situations the
renormalization procedure would be
different from the one described in this section due to the different
nature of the quantities $\Obs$. However, the general procedure of
Section~\ref{sec:renfree} still applies.

\subsection*{Acknowledgements}
This work was supported through a Marie Curie Fellowship of the
European Union.

\bibliography{stdrefs}
\bibliographystyle{amsordx}

\end{document}